\documentstyle[11pt,newpasp,twoside,epsf]{article}
\def\edcomment#1{\iffalse\marginpar{\raggedright\sl#1\/}\else\relax\fi}
\reversemarginpar

\begin{document}
\title{Approaching NGC\,3079 with VLBI}
 \author{E. Middelberg, T. P. Krichbaum, A. L. Roy, A. Witzel, J. A. Zensus}
\affil{Max-Planck-Institut f\"ur Radioastronomie, Bonn, Germany}

\begin{abstract}

The Seyfert 2 galaxy NGC\,3079 is one of the nearest radio-weak
AGN. We have carried out multi-frequency, multi-epoch VLBI
observations of this object and detected new components, proper motion
of the order of $0.1\,c$, strong spectral variability, and free-free
absorption. Based on phase-referenced astrometry, we suggest a new
identification of components seen at 1.7\,GHz. Our results suggest
that the nucleus in NGC\,3079 is embedded in a dense medium which
strongly affects the radio appearance of the AGN ejecta. We supplement
our work with Seyfert VLBI observations taken from the literature.
\end{abstract}

\section{Introduction}

Structure and kinematics in the jets of radio-loud AGNs have been
studied for long time spans, yielding discoveries such as superluminal
motion, flux density outbursts which mark the ejection of new jet
components, and twisted magnetic fields within the jets. Only little
is known about the physically less extreme radio-weak objects, and
only those few that are strong enough are suitable for detailed VLBI
studies. NGC\,3079 is one of the brightest Seyfert 2 galaxies, and
shows significant changes in radio configuration, component spectra
and overall geometry on a timescale of a few years. It can contribute
important insights to our understanding of the difference between
radio-loud and radio-weak AGN.

\section{NGC 3079}

NGC~3079 is a nearby Seyfert 2 or LINER galaxy at a distance of 15~Mpc
with embedded ${\rm H_2O}$ maser emission. VLA B-array observations at
6~cm show a bright core and two prominent radio lobes. These lobes
extend 2\,kpc away from the nucleus in P.A. $70^{\circ}$, almost
perpendicular to the disc of the galaxy in P.A. $165^{\circ}$ (Duric
et al. 1983). Pilot VLBI observations by Irwin \& Seaquist (1988)
resolved the core into two strong components, $A$ and $B$, separated
by 1.47\,pc in P.A. $130^{\circ}$ (relative to $A$). They also found
emission along a line between these components, suggestive of a
jet-like feature, which they named $C$. These three components do not
align with either the kpc-scale lobes or the galaxy disc.

Spectral-line and continuum observations by Trotter et al. (1998) and
Sawada-Satoh et al. (2000) (hereafter S00) have resolved the ${\rm
H_2O}$ maser emission in NGC~3079 into numerous aligned spots at
almost the same P.A. as the galaxy disc and at an angle of
$\approx35^{\circ}$ with respect to the jet axis. Those authors
suggest that the location of the nucleus is on the $A$-$B$ axis, where
it intersects with the line of masers, although there is no radio
emission from that location. Another component, $D$, was found by T98
further down the $A$-$B$ axis but was not confirmed by S00. $A$ and
$B$ were found by S00 to separate at $0.16\,c$, and $B$ appeared to be
stationary with respect to the brightest maser clumps.

\section{Observations and Results}

We have observed NGC\,3079 at 1.7\,GHz, 2.3\,GHz, 5.0\,GHz and
15.4\,GHz during four epochs (Table~1). All observations were made
with the VLBA and Effelsberg\footnote{The National Radio Astronomy
Observatory is a facility of the National Science Foundation, operated
under cooperative agreement by Associated Universities, Inc. The
Effelsberg 100-m telescope is operated by the Max-Planck-Institut
f\"ur Radioastronomie.}, except for the 2.3 and 5.0\,GHz observations
in 2002-09-22 which were made with the VLBA only. Only the
observations of the last epoch included polarization measurements and
were phase-referenced to a calibrator. At 1.7\,GHz the source is very
weak and resolved and was not detected on baselines longer than
$5\,{\rm M\lambda}$. Phase-referencing, however, yielded a good (${\rm
SNR}=8$) detection. The determination of instrumental polarization did
not work at 2.3\,GHz, and so polarization information could be
obtained only at 1.7\,GHz and 5.0\,GHz. At 15.4\,GHz, NGC\,3079 was
successfully detected on baselines shorter than $200\,{\rm M\lambda}$.

\begin{table*}[thpb]
\scriptsize
\begin{center}
\begin{tabular}{lc|rrrrrr|rrr}
\hline
\hline
Epoch        &  \multicolumn{1}{c}{Freq.} & \multicolumn{1}{|c}{Peak} & \multicolumn{1}{c}{rms} & \multicolumn{4}{c|}{Int. flux / mJy} & \multicolumn{3}{c}{Separations / mas}\\
             &  \multicolumn{1}{c}{GHz}   & \multicolumn{2}{|c}{mJy\,beam$^{-1}$} & \multicolumn{1}{c}{$A$} & \multicolumn{1}{c}{$B$} & \multicolumn{1}{c}{$E$} & \multicolumn{1}{c|}{$F$} & \multicolumn{1}{c}{$A-B$} & \multicolumn{1}{c}{$B-E$} & \multicolumn{1}{c}{$A-E$}\\
\hline
1999-11-20 &   5.0  & 20   &  0.08  & 27    & 17    &  4.2  &      & 26.2  & 26.3  & 17.7  \\ 
       	   &  15.4  & 45   &  0.33  &  5.7  & 47    &       &      &                       \\ 
2000-03-06 &   5.0  & 17   &  0.06  & 25    & 18    &  3.1  &      & 26.4  & 26.2  & 17.7  \\ 
       	   &  15.4  & 51   &  0.35  &  5.9  & 54    &       &      &                       \\ 
2000-11-30 &   5.0  & 18   &  0.06  & 24    & 18    &  5.2  &      & 26.5  & 25.9  & 17.6  \\ 
       	   &  15.4  & 51   &  0.23  & 17    & 62    &       &      &                       \\     
2002-09-22 &  1.7   & 9.1  &  0.06  &       &       &  3.1  & 16.7 &                       \\ 
       	   &  2.3   & 20   &  0.13  &  0.57 &       & 18    & 2.6  &                       \\ 
       	   &  5.0   & 22   &  0.11  & 25    & 21    &  7.1  &      & 27.2  & 25.5  & 17.7  \\ 
\hline
\end{tabular}
\caption{ Image and component parameters. Integrated flux
density errors are typically 5\,\% and separation errors are typically
0.20\,mas (1\,$\sigma$).}
\end{center}
\end{table*}

\begin{figure}
\plotone{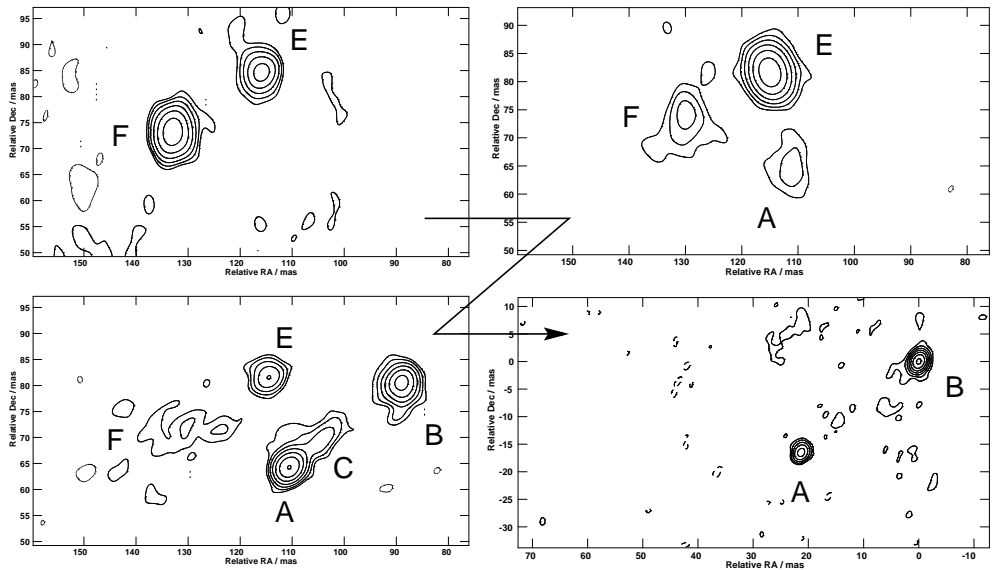}
\caption{NGC\,3079 VLBI images at 1.7\,GHz (upper left), 2.3\,GHz (upper
right), 5.0\,GHz (lower left) and 15.4\,GHz (lower right). The arrow
denotes increasing frequency.}
\end{figure}

The images from 2002-09-22 and the 15.4\,GHz image from 2000-11-30 are
shown in Fig.~1. We have detected two new components $E$ and $F$ at
1.7\,GHz to 5.0\,GHz in addition to the known components $A$, $B$ and
$C$. Two components are visible at 1.7\,GHz, a more compact one to the
north-west ($E$) and an extended one to the south-east ($F$). Based on
absolute coordinates, we find that these components are not coincident
with $A$ and $B$ seen at higher frequencies, despite having a similar
separation and relative position angle. They are, instead, coincident
with $E$ and $F$ seen at 5.0\,GHz. At 2.3\,GHz, component $F$ has
weakened, but $E$ has become stronger. South of $E$, component $A$
starts to show up, but $B$ is still not detected. Only at 5.0\,GHz are
all components visible, including the jet-like feature $C$ and a weak
detection of $F$.  At 15\,GHz, only $A$ and $B$ are detected. Our
polarimetric observations did not show polarized emission at 1.7\,GHz
or 5.0\,GHz, the limits being 2.2\,\% and 1.2\,\%, respectively.\\

\subsection{Spectra and Spectral Variability}

The big spectral differences among the components are illustrated in
Fig.~2 (left panel). The graphs show the spectra of (top to bottom)
$F$, $E$, $A$ and $B$. 

Component $A$ has a highly inverted spectrum with
$\alpha^{5.0}_{2.3}=4.85$ between 2.3\,GHz and 5.0\,GHz and a steep
spectrum with $\alpha^{5.0}_{15.4}=-0.28$ (2000-11-30). However, this
has significantly evolved from $\alpha^{5.0}_{15.4}=-1.38$ in
1999-11-20. Component $B$ also has an inverted spectrum with
$\alpha^{5.0}_{15.4}=0.90$ to 1.08 during the first three epochs. $B$
was detected only at 5.0\,GHz in 2002-09-22, yielding a lower limit
of $\alpha^{5.0}_{2.3}>4.5$.

\begin{figure}[htpb!]
\plottwo{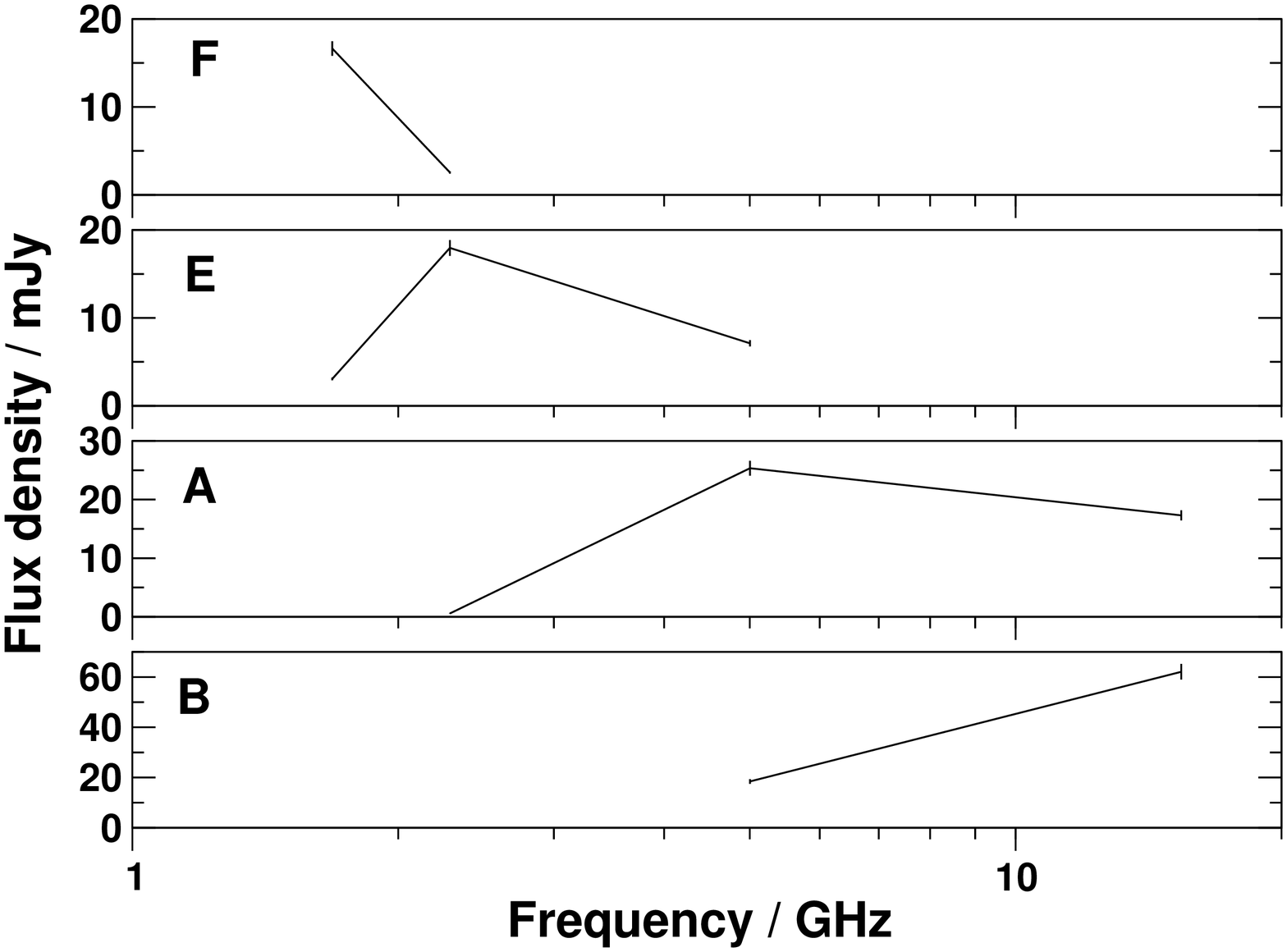}{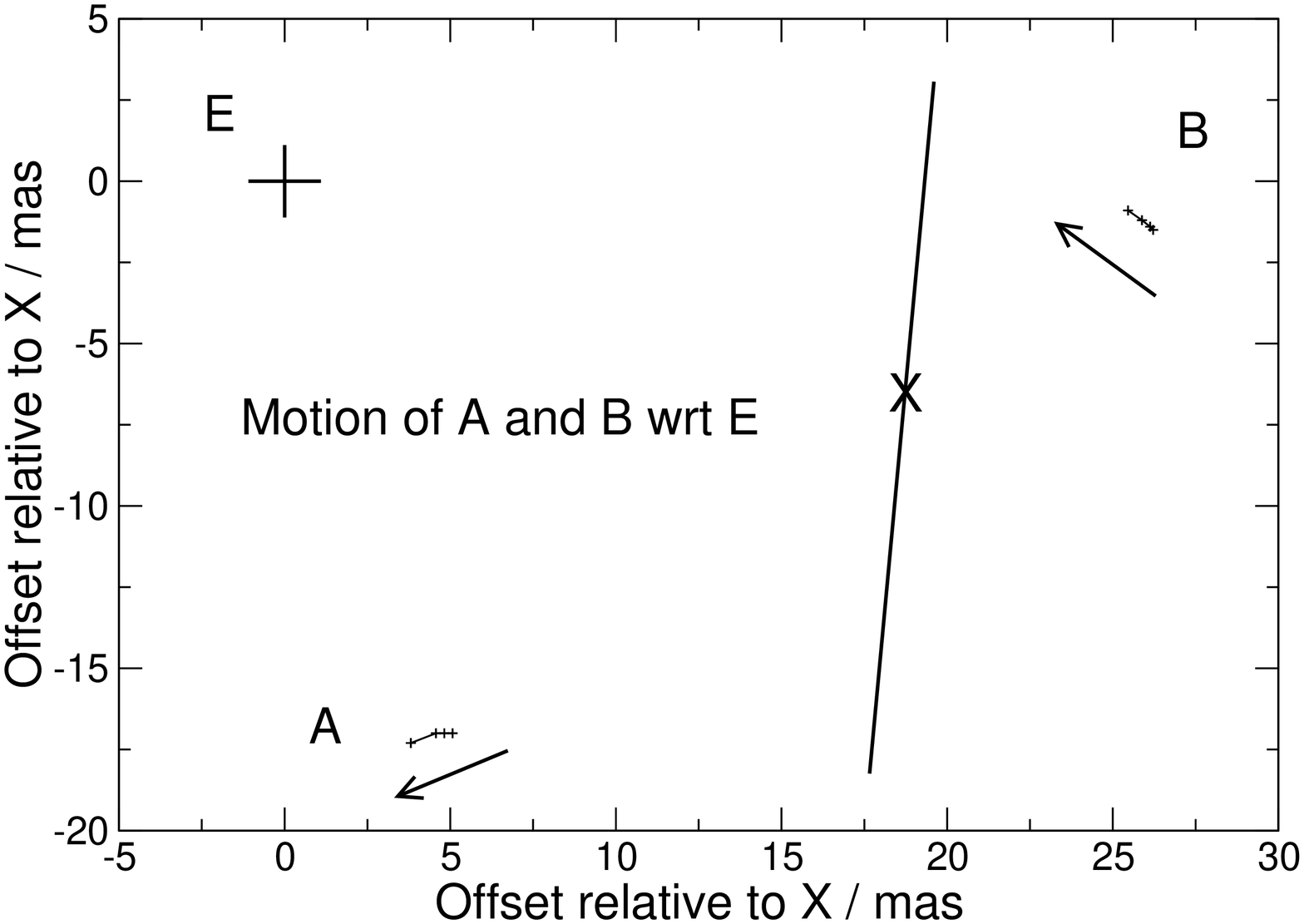}
\caption{{\it Left panel:} Spectra of components $F$, $E$, $A$ and $B$
(top to bottom). {\it Right panel:} Positions of $A$ and $B$ relative
to $E$. The vertical line indicates the projection of the disc of
masers, the ``X'' denotes the intersection of the maser disc with
the $A-B$ axis and the arrows indicate the directions of motion.}
\end{figure}

The new component $E$ has a very peculiar, narrow-peaked spectrum,
with $\alpha^{1.7}_{2.3}=5.77$ and
$\alpha^{2.3}_{5.0}=-1.18$. Synchrotron self-absorption cannot explain
the inverted side with $\alpha^{1.7}_{2.3}=5.77$, and free-free
absorption (FFA) in a foreground absorber is the most likely cause. If
FFA is present, the observed spectrum will be $S_{\nu}^{\rm
FFA}=S_0\,\nu^{\alpha_{\rm intr}}\times\,e^{-\tau_{\rm FFA}}$, where
$\tau_{\rm FFA}\propto\nu^{-2.1}$, resulting in an exponential cutoff
towards lower frequencies. However, fitting a spectrum of this form to
the data yields an uncommonly {\it steep} intrinsic spectral index
(before the absorber) of $\alpha_{\rm intr}=-4.59$. A possible
explanation could be that the electrons in the source population are
not replenished and age, causing a cutoff towards high frequencies. If
so, the FFA and electron population cutoffs would need to be very
close in frequency. The spectrum of $F$ is extremely steep between
1.7\,GHz and 2.3\,GHz, with $\alpha^{1.7}_{2.3}=-6.12$. We consider it
most unlikely that the extreme spectra of $E$ and $F$ are due to
resolution effects, because we used approximately scaled arrays (i.e.,
Effelsberg was included at 1.7\,GHz only) and the data were calibrated
and analyzed in the same manner, using the same tapering, weights, and
imaging procedures.

The peak frequencies increase with decreasing RA of the components,
suggestive of a gradient in a foreground absorber (in the shape of a
wedge?) or, alternatively, an absorber with considerable and clumpy
sub-structure on sub-pc scales.

\subsection{Proper Motions}

As absolute positions are lost in phase self-calibration, the analysis
of motions inside NGC\,3079 is restricted to relative position
changes. The component separations shown in Table~2 yield speeds of
$v_{A-B}=(0.11\pm0.01)\,c$, $v_{B-E}=(-0.07\pm0.01)\,c$ and
$v_{A-E}=(0.00\pm0.01)\,c$, where the determination of $v_{A-B}$ has
been supplemented by literature data. Selecting either $A$, $B$ or $E$
as the stationary reference point then yields three different
scenarios, and we consider the one shown in Fig.~2 ($E$ stationary)
the most natural.  Although we do not believe that $E$ is the nucleus,
it might be kinematically disconnected from the AGN and hence be
quasi-stationary over the observed epochs. The back-extrapolations of
the trajectories of $A$ and $B$ then intersect in the past, from which
point they may have been ejected. However, the nature of $E$ remains
unclear. It cannot be a starburst due to its high brightness
temperature, $T_{\rm B}=4.6\,\times10^8\,{\rm K}$ at 2.3\,GHz, but it
could be a supernova. In previously suggested scenarios (S00,
Greenhill et al. 2003, this conference), $B$ was assumed to be
stationary. If so, $E$ would be moving {\it inwards}, perpendicular to
the $A$-$B$ axis, at $0.08\,c$. Although the evidence from the maser
observations is strong, we consider this situation more difficult to
understand than the scenario in which $E$ is stationary.

\section{Statistics}

\begin{figure}
\plottwo{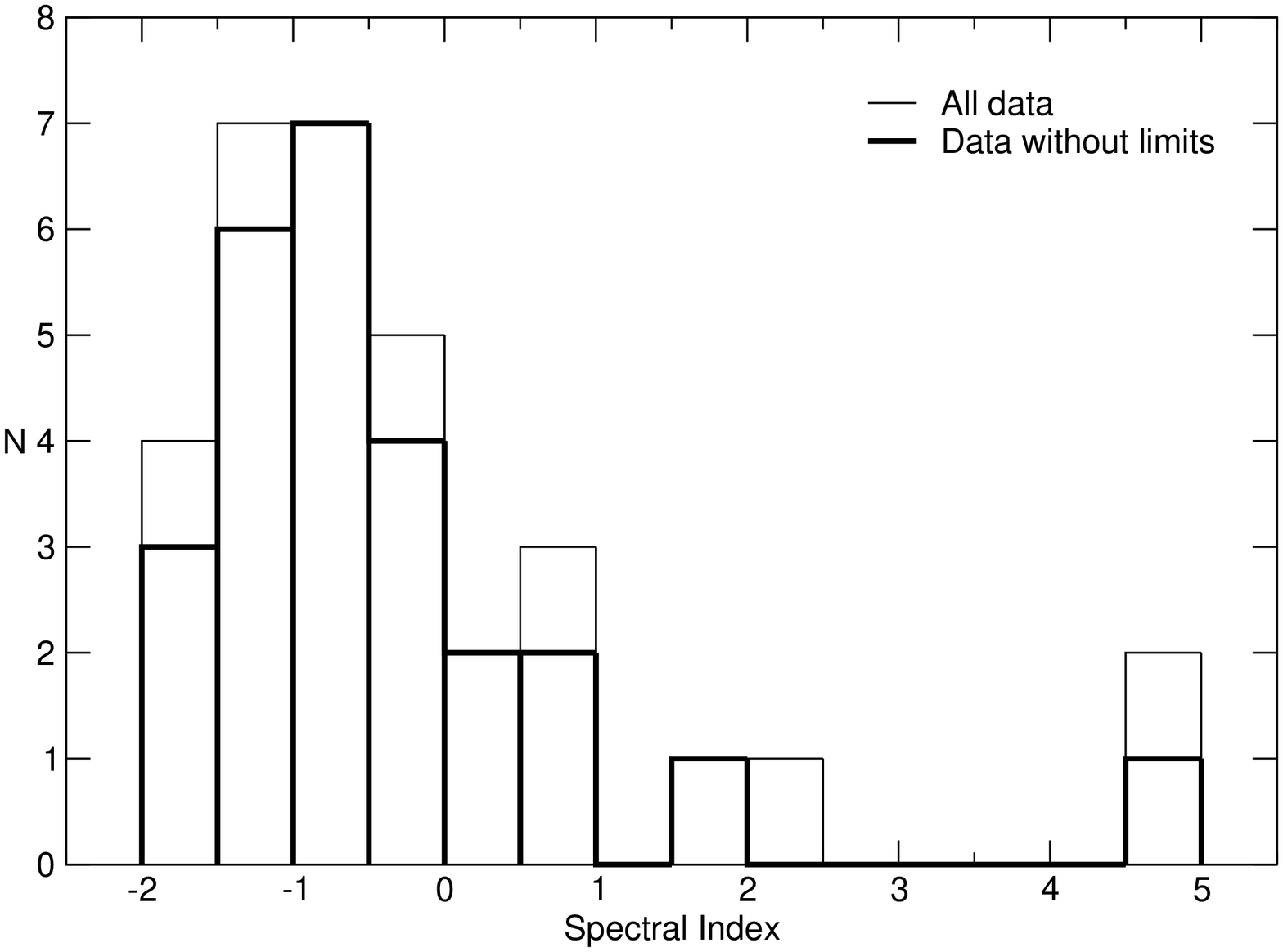}{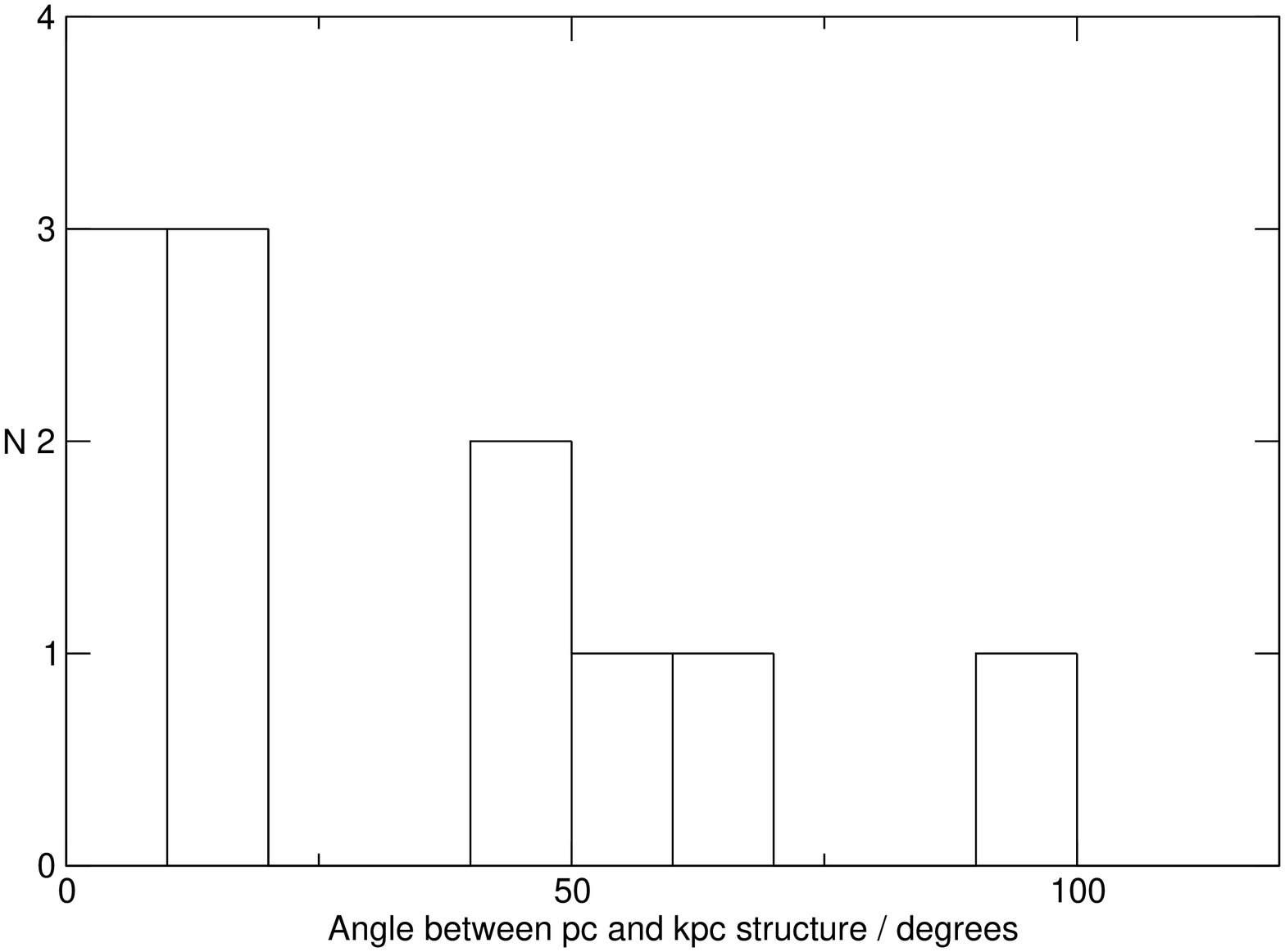}
\caption{{\it Left panel:} Histogram of component spectra in Seyfert
galaxies measured with VLBI. {\it Right panel:} Histogram of angles
between kpc and pc-scale structure in Seyfert galaxies}
\end{figure}

We have searched the literature for VLBI observations of Seyfert
galaxies to see how NGC\,3079 compares to other objects. We have found
33 spectral indices of components in 13 Seyferts, and 11 Seyferts
where a linear structure is seen both on pc and kpc scales (Fig.~3).
The spectral indices cluster around $-0.75$, as expected for optically
thin synchrotron emission, and then fade into a tail out to +5. The
points between $\alpha=4.5$ and $\alpha=5$ were measured in NGC\,3079
(components $A$ and $B$), indicating that this galaxy is a special
object. However, 10 out of 13 objects have flat ($-0.3<\alpha<0.3$) or
inverted ($\alpha\ge0.3$) spectra, indicating compact synchrotron
emission or free-free absorption.

The pc- and kpc-scale structures in Seyferts show no strong tendency
to be aligned: 8 out of 11 are misaligned with angles of $>15^{\circ}$
between pc and kpc-scale structure. This indicates a difference from
powerful radio objects, which show a bimodal distribution, tending to
be either aligned or perpendicular (Pearson \& Readhead 1988, Conway
\& Murphy 1993). The frequent misalignment in Seyferts indicates
either precession of the jet ejection axis or pressure gradients and
interactions with the ISM on scales of tens to hundreds of pc.

\end{document}